\documentclass[useAMS]{mn2e} 
\usepackage{times} 
\usepackage{epsfig} 
 
\newcommand{\etal} 
	{et al.} 
\newcommand{\eg} 
	{e.g.} 
 
\newcommand{\ct} 
	{\mu} 
 
\newcommand{\xir} 
	{\xi(r)}
\newcommand{\xisp} 
	{\xi(\sigma,\pi)}
\newcommand{\bj} 
        {b_{\rm J}} 
\def\kms{\;{\rm km\,s}^{-1}} 
\def\sigeight{\sigma_8^{\scriptscriptstyle\rm NL}} 
\def\hmpc{\,h^{-1}\,{\rm Mpc}}
 
\begin{document} 
 
\title[The 2dFGRS: galaxy clustering per spectral type]      
{The 2dF Galaxy Redshift Survey: galaxy clustering 
per spectral type}            
 
\author[D.S. Madgwick \etal]{    
       \parbox[t]{\textwidth}{Darren S.\ Madgwick$^{1,2}$, 
Ed Hawkins$^{3}$, 
Ofer Lahav$^{2}$, 
Steve Maddox$^{3}$, 
Peder Norberg$^{4}$, 
John A.\ Peacock$^{5}$, 
Ivan K.\ Baldry$^{6}$, 
Carlton M.\ Baugh$^7$, 
Joss Bland-Hawthorn$^8$, 
Terry Bridges$^8$,  
Russell Cannon$^8$,  
Shaun Cole$^7$,  
Matthew Colless$^9$,  
Chris Collins$^{10}$,  
Warrick Couch$^{11}$,  
Gavin Dalton$^{12,13}$, 
Roberto De Propris$^{11}$, 
Simon P.\ Driver$^9$,  
George Efstathiou$^2$,  
Richard S.\ Ellis$^{14}$,  
Carlos S.\ Frenk$^7$,  
Karl Glazebrook$^{6}$,  
Carole Jackson$^9$, 
Ian Lewis$^{12}$,  
Stuart Lumsden$^{15}$,  
Bruce A.\ Peterson$^9$,  
Will Sutherland$^{5}$, 
Keith Taylor$^{14}$ 
(The 2dFGRS Team)}
\vspace*{6pt} \\  
$^1$Department of Astronomy, University of California, Berkeley, CA
94720, USA \\
$^2$Institute of Astronomy, Madingley Road, Cambridge 
CB3 0HA, U.K. \\ 
$^{3}$School of Physics \& Astronomy, University of Nottingham, 
       Nottingham NG7 2RD, UK \\ 
$^{4}$Institut f\"ur Astronomie, ETH H\"onggerberg, CH-8093 Z\"urich,
Switzerland 
$^{5}$Institute for Astronomy, University of Edinburgh, Royal Observatory,  
       Blackford Hill, Edinburgh EH9 3HJ, UK \\ 
$^6$Department of Physics \& Astronomy, Johns Hopkins University, 
       Baltimore, MD 21218-2686, USA \\ 
$^7$Department of Physics, University of Durham, South Road,  
    Durham DH1 3LE, UK \\  
$^8$Anglo-Australian Observatory, P.O.\ Box 296, Epping, NSW 2121, 
    Australia\\   
$^9$Research School of Astronomy \& Astrophysics, The Australian  
    National University, Weston Creek, ACT 2611, Australia \\ 
$^{10}$Astrophysics Research Institute, Liverpool John Moores University,   
    Twelve Quays House, Birkenhead, L14 1LD, UK \\ 
$^{11}$Department of Astrophysics, University of New South Wales, Sydney,  
    NSW 2052, Australia \\ 
$^{12}$Department of Physics, University of Oxford, Keble Road,  
    Oxford OX1 3RH, UK \\ 
$^{13}$Space Science and Technology Division, Rutherford Appleton 
Laboratory, Chilton, Didcot, OX11 0QX, UK \\
$^{14}$Department of Astronomy, California Institute of Technology,  
    Pasadena, CA 91125, USA \\ 
$^{15}$Department of Physics, University of Leeds, Woodhouse Lane, 
       Leeds, LS2 9JT, UK \\ 
}

\date{ 
MNRAS submitted - 2003 March 29 
} 
  
\pagerange{\pageref{firstpage}--\pageref{lastpage}} 
\pubyear{2003} 
 
\label{firstpage} 
 
\maketitle 
 
\begin{abstract} 
We have calculated the two-point correlation functions in redshift
space, $\xi(\sigma,\pi)$, for galaxies of different spectral types in
the 2dF Galaxy Redshift Survey.  Using these correlation functions we
are able to estimate values of the linear redshift-space distortion
parameter, $\beta\equiv\Omega_{\rm m}^{0.6}/b$, the pairwise velocity
dispersion, $a$, and the real-space correlation function, $\xi(r)$,
for galaxies with both relatively low star-formation rates (for which
the present rate of star formation is less than 10\% of its past
averaged value) and galaxies with higher current star-formation
activity.  At small separations, the real-space clustering of passive
galaxies is very much stronger than that of the more actively
star-forming galaxies; the correlation-function slopes are
respectively 1.93 and 1.50, and the relative bias between the two
classes is a declining function of radius. On scales larger than 
$10 h^{-1}$ Mpc
there is evidence that the relative bias tends to a constant,
$b_{\rm passive}/b_{\rm active} \simeq 1$.
This result is consistent with the similar degrees of
redshift-space distortions seen in the correlation functions of the
two classes -- the contours of $\xi(\sigma,\pi)$ 
require $\beta_{\rm active}=0.49\pm0.13$, and 
$\beta_{\rm passive}=0.48\pm0.14$.  
The pairwise velocity dispersion is highly correlated with $\beta$.
However, despite this a significant
difference is seen between the two
classes.  Over the range $8-20\;h^{-1}$~Mpc, the pairwise velocity
dispersion  has mean values
$416\pm76\kms$ and $612\pm92\kms$ for the active and passive galaxy
samples 
respectively.  This is consistent with the expectation from morphological
segregation, in which passively evolving galaxies preferentially
inhabit the cores of high-mass virialised regions.
\end{abstract} 
 

 
\section{Introduction} 
\label{section:intro} 
 
It is now well established that the clustering of galaxies at low
redshift depends on a variety of factors.  Two of the most prominent
of these, which have been discussed extensively in the literature, are
luminosity (see e.g. Norberg \etal~2001 and references therein) and
galaxy type (e.g. Davis \& Geller 1976; Dressler 1980; Lahav, Nemiroff
\& Piran 1990; Loveday
\etal~1995; Hermit \etal~1996; Loveday, Tresse \& Maddox 1999; Norberg
\etal\ 2002a).  It is the latter of these which we wish to address in
this paper, by making use of the 221,000 galaxies observed in the
completed 2dF Galaxy Redshift Survey (2dFGRS; Colless \etal~2001).
 
Previous analyses of the clustering of galaxies as a function of
morphological type have revealed that early-type galaxies are
generally more strongly clustered than their late-type counterparts;
their correlation function amplitudes being up to several times
greater (e.g. Hermit \etal~1996).  These results are also found to
hold true if one separates galaxies by colour (Zehavi \etal~2001) or
spectral type (Loveday, Tresse \& Maddox 1999), both of which are
intimately related to the galaxy morphology (see e.g. Kennicutt~1992).
 
The existence of this distinction in the clustering behaviour of
different types of galaxies is to be expected if one considers
galaxies to be biased tracers of the underlying mass distribution,
since the amount of biasing should be related to a galaxy's mass and
formation history. However, the fact that the most recent analyses of
the total galaxy population have revealed that galaxies are not on
average strongly biased tracers of mass on large scales (Lahav \etal\
2002; Verde \etal\ 2002) makes this behaviour even more interesting,
and puts some degree of perspective on these results.
 
In this paper we attempt to make the most accurate measurements of 
this distinctive clustering behaviour, by calculating the 
two-dimensional correlation function, $\xi(\sigma,\pi)$, for the 
most quiescent and star-forming galaxies in our sample separately, 
where $\sigma$ is the galaxy separation perpendicular to the 
line--of--sight and $\pi$ parallel to the line--of--sight.  This 
simple statistic allows us to easily visualise and quantify the variation in 
clustering properties on a variety of scales, picking out for example 
the distinctive `finger-of-God' effect due to peculiar velocity 
dispersions in virialised regions, as well as the large scale 
flattening due to coherent inflows of galaxies towards over-dense 
regions. 
 
By contrasting these observed effects we can gain significant insights 
into the properties of the galaxy population, particularly when these 
results are set against simple analytic models.  For example, we can 
use the large--scale inflows to 
constrain the quantity $\beta\equiv \Omega^{0.6}/b$, and the 
small--scale `finger-of-God' distortions to constrain the distribution 
of galaxy peculiar velocities $f(v)$ simultaneously.  Such an analysis 
has already been performed using an earlier subset of the 2dFGRS 
by Peacock 
\etal\ (2001), and an updated version of that analysis is presented in 
Hawkins \etal\ (2003).  This paper extends their analyses by 
incorporating the spectral classification of 2dFGRS galaxies presented 
in Madgwick \etal\ (2002a). 
 
The outline of this paper is as follows.  In Section~2 we briefly describe 
the 2dFGRS data-set and the spectral classification we are adopting.  
In Section~3 we then outline the methods we use for estimating the 
correlation function and the models we use when making fits to this function.
The results of our parameter fits are presented in Section~4 and are compared
to previous results in Section~5.  In Section~6 we conclude this paper with a 
discussion of our results.

\section{The 2\lowercase{d}FGRS Data} 
 
The data-set used in this analysis consists of a subset of that 
presented in Hawkins \etal\ (2003) -- including only those galaxies with 
spectral types (Section~\ref{section:section1}), 
which lie in the redshift interval $0.01<z<0.15$. 
Again we restrict ourselves to only considering the most complete 
sectors of the survey, for which $>70$\% of the galaxies have 
successfully received redshifts.   
This leaves us with 96\,791 galaxies for use in the analysis presented 
here.  Further details of the data-set are presented in Hawkins \etal\ 
(2003).

\subsection{Spectral types} 
\label{section:section1} 
 
We adopt here the spectral classification developed for the 2dFGRS in 
Madgwick \etal\ (2002a).  This classification, $\eta$, is derived from 
a principal components analysis (PCA) of the galaxy spectra, and 
provides a continuous parameterisation of the spectral type of a 
galaxy based upon the strength of nebular emission present in its 
rest-frame optical spectrum.   
It is found that $\eta$ correlates 
relatively well with galaxy $B$-band morphology (Madgwick 2002).  
However, the most natural interpretation of 
$\eta$ is in terms of the relative amount of star formation 
occurring in each galaxy, parameterised in terms of the Scalo 
birthrate parameter, $b_{\rm Scalo}$ (Scalo 1986), 
\begin{equation} 
b_{\rm Scalo} = \frac{SFR_{\rm present}}{\langle SFR \rangle_{\rm past}} 
\end{equation} 
as demonstrated in Madgwick \etal\ (2002b). 
 
Although $\eta$ is a continuous variable we find it convenient to 
divide our sample of galaxies at $\eta=-1.4$. 
It is found that this cut of $\eta=-1.4$ corresponds 
to approximately $b_{\rm Scalo}=0.1$ 
(i.e. the current rate of star formation is 10\% of its past 
averaged value).
 
The cut in $\eta$ we have adopted is the same as that used to
distinguish the so-called `Type 1' galaxies used in our calculation of
the galaxy $\bj$ luminosity functions (Madgwick \etal\ 2002a).  In
that paper two more cuts were made at $\eta=1.1$ and $\eta=3.5$, which
we have not adopted in this analysis.  It is found that using only two
spectral types instead of four greatly increases the accuracy of our
analysis, while the clustering properties of the most actively
star-forming galaxies are found to be very similar
(Section~\ref{section:xir}).  The clustering with spectral type in the
2dFGRS has previously been considered by Norberg \etal\ (2002a);
however the present paper extends this analysis by considering the
full magnitude limited survey.
 
In the analysis that follows the two samples constructed by dividing
at $\eta=-1.4$ will be referred to as the {\em relatively} passive and
active star-forming galaxy samples.  These two samples consist of a
total of 36318 and 60473 galaxies respectively.

\section{The two-point correlation function} 
\label{section:section2} 

\begin{figure*} 
\psfig{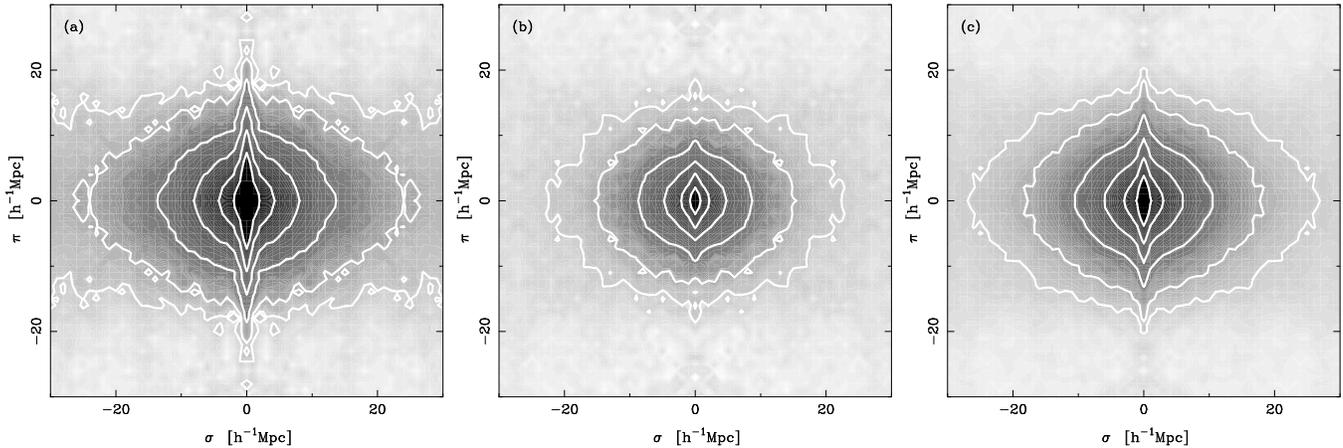}
\caption{The $\xi(\sigma,\pi)$ grids for our different spectral
types: (a) passive, (b) active and (c) full samples. 
The contour levels are at $\xi$
= 4.0, 2.0, 1.0, 0.5, 0.2 and 0.1.}
\label{fig:sp}
\end{figure*}

The correlation function, $\xi$, is measured by comparing the actual 
galaxy distribution with a catalogue of randomly distributed 
galaxies.  These randomly distributed galaxies are subject to the same 
redshift and mask constraints as the real data.  $\xi(\sigma,\pi)$ is 
estimated by counting the pairs of galaxies in bins of separation 
along the line-of-sight, $\pi$, and across the line-of-sight, 
$\sigma$, using the following estimator, 
\begin{equation} 
\xi(\sigma,\pi) = \frac{\langle DD\rangle -2\langle DR\rangle -\langle 
RR\rangle}{\langle RR\rangle }\;, 
\end{equation} 
from Landy \& Szalay (1993).  In this equation $\langle DD\rangle$ is 
the weighted number of  
galaxy-galaxy pairs with particular ($\sigma$,$\pi$) separation, 
$\langle RR\rangle$  
the number of random-random pairs and $\langle DR\rangle$ the number of 
data-random pairs.  The normalisation adopted is that the sum of
weights for the real galaxy catalogue should match that of the random
catalogue as a function of scale.  As in Hawkins \etal~(2003) we 
adopt the $J_3$ weighting scheme to minimise the variance in the 
estimated correlation function 
(Peebles 1980).  We have also estimated the correlation functions
using the estimator of Hamilton (1993), however because these give
an essentially identical estimate for $\xi$ (well within the statistical
uncertainties) we only present the 
results from the Landy \& Szalay estimator in this paper.

The random catalogue is constructed by generating random positions on 
the sky and modulating the surface density of these points by the 
completeness variations of the 2dFGRS.  Note that, in contrast to
Hawkins et al. (2003), this completeness now
also includes the that introduced by the spectral classification for
galaxies with $z<0.15$ (Norberg 2001).  The redshift distribution is 
then drawn from the selection function of each type as calculated from 
the 2dFGRS luminosity functions -- which allow us to naturally
incorporate the varying magnitude limit of the survey.  
These luminosity functions have been 
calculated as in Madgwick \etal~(2002a) for the data used in this 
analysis in both the NGP and SGP regions separately for each spectral 
type. 
 
Due to the design of the 2dF instrument, fibres cannot be placed
closer than approximately 30 arcsec (Lewis \etal~2002). In Hawkins
\etal~(2003) the effects of these so-called `fibre collisions' were
taken into account in the estimation of the correlation functions by
comparing the angular correlation functions of the parent and redshift
catalogues. It was found that the effect was significant for
separations $< 0.2 \hmpc$. This cannot be done for the present
analysis because of the spectral type selection and so we ignore all
separations $< 0.2 \hmpc$ in the fitting process.
 
The resulting estimates of $\xi(\sigma, \pi)$ are presented in
Fig.~\ref{fig:sp} and clear differences are immediately visible. The
rest of this paper is spent quantifying these differences.

\subsection{De-correlating error bars} 
\label{section:decorrelation} 
 
Many of the subsequent sections of this paper will involve attempting
to fit a parametric form to $\xi(\sigma,\pi)$ or $\xi(r)$.  Because
the individual points we estimate for each of these quantities and
their associated error bars are not independent (a single galaxy can
contribute to correlations over all scales), a simple $\chi^2$ fit
between the observed and model correlation functions may not yield the
most accurate parameter fit.  For this reason we must carefully
account for these correlations in each of our subsequent fits.

When fitting the model correlation function to the observed 
$\xi$, we are interested in minimising the residual 
between the two.  For this reason we make a simple change of variables 
to define, 
\begin{equation} 
\Delta(s_i) = \frac{(\xi(s_i) - \langle \xi(s_i) \rangle)}{\sigma_{\xi}(s_i)}\;. 
\end{equation} 
Where here $\xi(s)$ is a given realisation of the correlation 
function we are estimating, 
$\langle \xi(s_i) \rangle$ is the mean value of our ensemble of
correlation functions at separation $s_i$, 
and $\sigma_{\xi}(s_i)$ is the standard deviation of the estimates of 
the correlation function at this same separation, as deduced from the
bootstrap analysis described below. 
We then construct the covariance matrix, 
\begin{equation} 
C_{ij} = \langle \Delta(s_i) \Delta(s_j) \rangle 
\end{equation} 
in terms of these variables. 
  
The best-fitting model correlation function 
$\xi^{\rm model}({{s}})$ can be found through  minimising 
the residual between it and the observed correlation function 
$\xi^{\rm obs}({{s}})$ in terms of the $\chi^2$ difference between the 
two.  The residual between the models and observations is defined by, 
\begin{equation} 
\Delta^{\rm (res)}(s_i) = \frac{\bigl(\xi^{\rm obs}(s_i) - \xi^{\rm model}(s_i)\bigr)} 
{\sigma_\xi(s_i)}\;, 
\label{eqn:vector} 
\end{equation} 
in which case the $\chi^2$ can be found from, 
\begin{equation} 
\chi^2 = (\vec{\Delta}^{\rm (res)})^{\rm T} C^{-1} \vec{\Delta}^{\rm (res)} \;. 
\end{equation} 
Where $\vec{\Delta}^{\rm (res)}$ is the vector of elements given in 
Eqn.~\ref{eqn:vector}.  The above equations can then easily be generalised 
for the gridded $\xi(\sigma,\pi)$. 
Because the points in the $\xi(\sigma,\pi)$ grid are highly 
correlated there are in fact very few independent components in the 
observed correlation function.  For this reason it is also possible to 
extend this analysis as demonstrated by Porciani \& 
Giavalisco (2001) by instead first performing a principal component 
analysis using our estimated covariance matrix, however we have not
found this step to be necessary for this analysis. 
 
One would require a set of
independent realisations in order to determine
unbiased estimates of $\sigma_\xi$ and the data covariance matrix $C$.  
However, this is of
course not possible since we have at our disposal only one realisation
of the Universe.  In most cases a good determination of this
covariance can nonetheless be determined from mock galaxy simulations
(e.g. Cole \etal\ 1998).  These simulations are very useful in that
they represent the total expected cosmic variance.

The complication we encounter as opposed to other analyses
(e.g. Hawkins et al. 2003) is that the simulations
available to us cannot adequately account for the variation in galaxy
clustering for different types of galaxies. 
For this reason these
simulations can only  provide us with rough estimates as to the
magnitude of the cosmic variance -- which we must somehow scale to
correspond to our results.

Another method for determining the expected covariance of our data is
to use a bootstrap re-sampling of the data-set (Ling, Barrow \& Frenk
1986).  The most important assumption in a bootstrap estimation of the
covariance matrix is that each of the data-points sampled must be
independent.  This is not true of the galaxy distribution itself, but
if we divide our survey area into a selection of contiguous regions
and re-sample these in our bootstrap calculations this assumption will
hold (so long as each of the sectors of sky is large enough to be
representative).  For this reason, in our subsequent analysis, we
divide the SGP region of the 2dFGRS survey into eight sectors and the
NGP into six.  The selection of the regions has been made to ensure a
statistically significant and roughly equal number of galaxies in each
sector.  These regions are then selected at random, with replacement,
as in the standard bootstrap analysis.
We make use of 20 bootstrap realisations in the analyses that follow,
and use these to estimate the covariance matrix for each of our fits.
The limitation of the bootstrap approach is that the samples
are drawn from the observed volume of space, and may not represent 
the entire cosmic scatter.
However, we find that error bars on parameters derived from the mocks
using the procedure of Hawkins et al. (2003) 
are in reasonable agreement with those 
derived from our bootstrap approach (see Table 2).

\subsection{The real-space correlation function} 
\label{section:xir} 
 
Because the various redshift distortions to the correlation function
only affect its measurement along the line-of-sight, it is possible to
make an estimate of the real-space correlation function $\xi(r)$ by
first projecting the two-dimensional correlation function,
$\xi(\sigma,\pi)$, onto the $\pi=0$ axis.  This projected correlation
function, $\Xi(\sigma)$, is given by,
\begin{equation} 
\Xi(\sigma) = 2\int_0^\infty \xi(\sigma,\pi) d\pi \;. 
\end{equation} 
In practice the upper limit of the integration is taken to be a large
finite separation for which the integral is found to converge.  We
find that limiting the integration to $\pi=70$ $h^{-1}$Mpc suffices
for the analysis presented here, providing us with stable projections
out to $\sigma=30\;h^{-1}$~Mpc.
 
The projected correlation function can then be written as an integral
over the real-space correlation function (Davis \& Peebles 1983),
\begin{equation} 
\frac{\Xi(\sigma)}{\sigma} = \frac{2}{\sigma} \int_{\sigma}^\infty 
\frac{r\xi(r)dr}{(r^2-\sigma^2)^{1/2}} \;. 
\label{eqn:xisig} 
\end{equation} 
If we assume a power law form; $\xi(r)=(r/r_0)^{-\gamma}$, we can 
solve this equation for the unknown parameters, 
\begin{equation} 
\frac{\Xi(\sigma)}{\sigma} = \left( \frac{r_0}{\sigma} \right)^{\gamma} 
\frac{\Gamma(\frac{1}{2}) \Gamma(\frac{\gamma-1}{2})}{\Gamma(\frac{\gamma}{2})} 
\;. 
\end{equation}  
 
\begin{figure} 
\begin{center} 
\epsfig{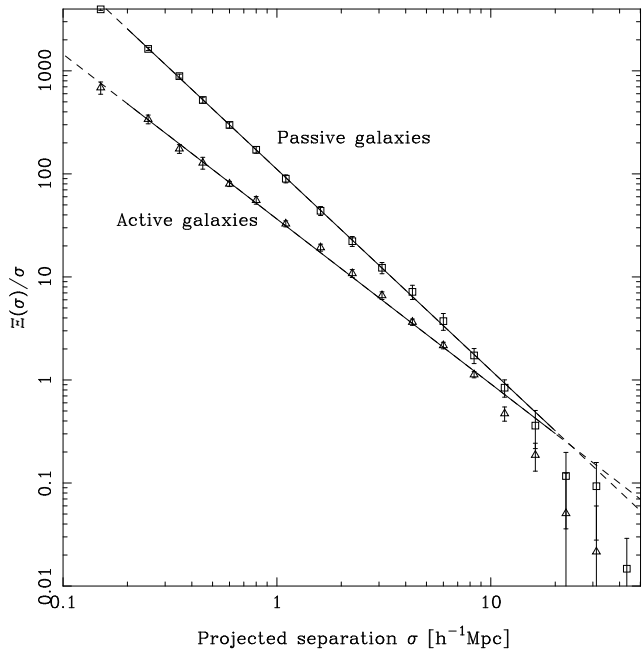} 
\caption[] {The projected
correlation function, $\Xi(\sigma)/\sigma$, is shown for both
relatively passive and active galaxies in the 2dFGRS.  It can be seen
that the correlation function of both sets of galaxies has an
approximate power law form for a large range of separations, and this
is illustrated with the best-fitting power law determined from this
data (solid line).  The dashed lines are extrapolations of these fits
to larger and smaller scales.}
\label{fig:Xi} 
\end{center} 
\end{figure}

The projected correlation functions, $\Xi(\sigma)/\sigma$, are shown in
Fig.~\ref{fig:Xi}, together with error bars derived from the bootstrap
realisations.  It can be seen that for both sets of galaxies the power
law assumption we have made is justified on small scales, and is quite
consistent with the observations on large scales.  The results from
fitting a power law form for $\xi(r)$ are given in Table~\ref{tabXi}
(Method $P$), together with those derived for the combined sample of
all spectrally typed galaxies.  Upon comparing with Hawkins \etal\
(2003), we find that our estimate of the combined correlation function
is essentially identical -- from which we can conclude that
restricting our analysis to only those galaxies with spectral types
has not biased our results in any noticeable way.
 
\begin{figure} 
\begin{center} 
\epsfig{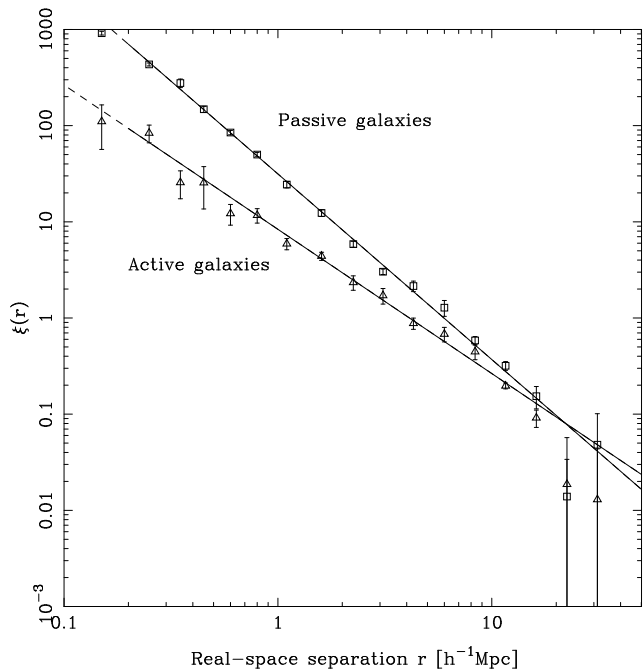} 
\caption[]{The non-parametric estimates of the real-space correlation
functions are shown for both our spectral types, using the method of
Saunders, Rowan-Robinson \& Lawrence (1992).  It can be seen that our
assumption of a power law form for $\xi(r)$ is justified out to scales
of up to $20\;h^{-1}$~Mpc.  The solid lines are the best-fitting power
law fits shown in Table~\ref{tabXi}, whereas the dashed lines are
extrapolations of these fits.} 
\label{fig:xx} 
\end{center} 
\end{figure}

In order to independently verify our assumption of a power law
$\xi(r)$, we have also calculated the real space correlation function,
using the non-parametric method of Saunders, Rowan-Robinson \&
Lawrence (1992) (Fig.~\ref{fig:xx}).  It can be seen that this method
also estimates a power-law form for $\xi(r)$ out to scales of $\sim
20\;h^{-1}$~Mpc and the best-fit parameters are shown in
Table~\ref{tabXi} (Method $I$).  We note however that a
clear shoulder appears to be present in both the correlation functions
for separations $r\sim8\hmpc$.  
It has recently been suggested that this
may reflect the transition scale between a regime dominated by galaxy
pairs in the same halo and a regime dominated by pairs in separate
halos (e.g. Zehavi \etal~2003; Magliochetti \& Porciani 2003).
 
To increase the accuracy of our results we have split the galaxy
sample into only two sub-samples. To justify this choice we have
repeated our analysis on two further sub-samples of the active
sample. We found that the $\xi(r)$ estimates were essentially
identical (and consistent within the estimated uncertainties), 
and so the clustering statistics are relatively insensitive to
the exact amount of star-formation occurring in active 
star-forming galaxies.

\begin{table} 
\begin{center}
\caption[] {The derived parameters
of the projected real-space correlation function.  The fits have been
determined using the range of separations $0.2 < r <
20\;h^{-1}$Mpc. The Method column refers to the projected ($P$) or
inverted ($I$) values of the parameters.  Also shown are the values of
$\sigeight$ derived from these correlation functions.  Note that the
lower limit of the fits ($0.2\;h^{-1}$~Mpc) was imposed to avoid
biases from fibre collisions (see Hawkins \etal\ 2003).}
 \begin{tabular}{@{}ccccc@{}}     
 \hline     
 Galaxy type & Method & $r_0$ ($h^{-1}$Mpc) & $\gamma$  & $\sigeight$ \\     
 \hline       
    All & $P$ & $4.69\pm0.22$ & $1.73\pm0.03$ & $0.83\pm0.06$ \\ 
   Passive & $P$ &$6.10\pm0.34$ & $1.95\pm0.03$ & $1.12\pm0.10$ \\ 
   Active & $P$ &$3.67\pm0.30$ & $1.60\pm0.04$ & $0.68\pm0.10$ \\     
\hline   
    All & $I$ & $5.01\pm0.23$ & $1.64\pm0.03$ & $0.88\pm0.05$ \\ 
   Passive & $I$ & $5.97\pm0.29$ & $1.93\pm0.03$ & $1.09\pm0.08$ \\ 
   Active & $I$ &$4.12\pm0.32$ & $1.50\pm0.04$ & $0.75\pm0.09$ \\    
\hline
\end{tabular}   
\label{tabXi}   
\end{center}  
\end{table}

\subsection{Relative bias} 
\label{section:relbias} 
 
The term {\em bias} is used to describe the fact that it is possible for 
the distribution of galaxies to not trace the underlying mass 
density distribution precisely.  The existence of such an effect would be a 
natural consequence if galaxy formation was enhanced, for example, in 
dense environments.  The simplest model commonly assumed (somewhat 
ad-hoc) to quantify the 
degree of biasing present, is that of the linear bias parameter, $b$, 
\begin{equation} 
\left( \frac{\delta\rho}{\rho}\right)_{\rm galaxies} = b \left( 
\frac{\delta\rho}{\rho} \right)_{\rm mass} \;, 
\end{equation} 
where $\rho$ is a measure of the density, of either the mass or the
galaxies.  A more specific model, based on the statistics of peaks
(Kaiser 1984; Bardeen \etal\ 1986), is that the degree of clustering
we observe in our galaxy sample, quantified in terms of the
correlation function, $\xi(r)$, is related to the mass correlation
function in terms of,
\begin{equation} 
\xi(r)_{\rm galaxies} = b^2 \xi(r)_{\rm mass} \;. 
\end{equation} 
Where here $b$ is a constant that does not vary with scale, but more 
generally may depend on $r$. 
 
\begin{figure} 
\begin{center} 
\epsfig{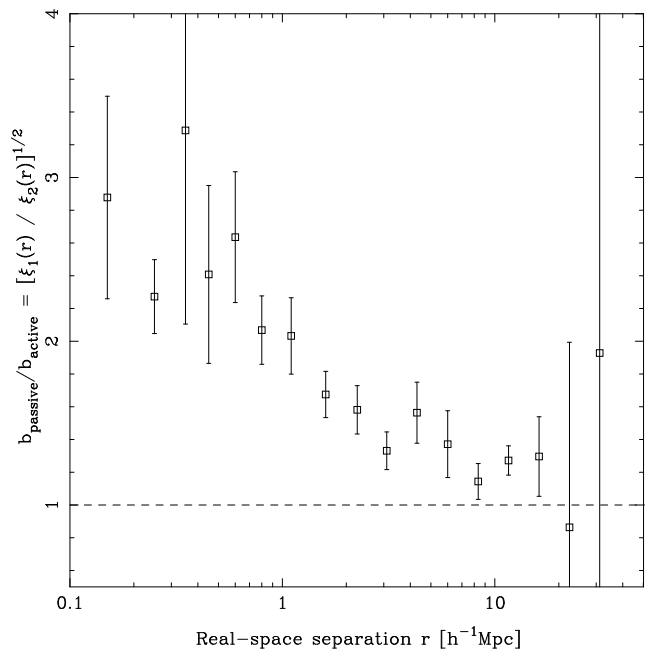} 
\caption[The relative biasing between spectral types] {The relative
bias between the most passive and actively star-forming galaxies is
shown, in terms of (the square-root of) the ratio of the real-space
correlation functions of these two samples.}
\label{fig:xibias} 
\end{center} 
\end{figure}

It is possible to estimate the magnitude of the biasing present in a sample 
of galaxies through the use of `redshift-space distortions', and this 
issue will be addressed later in this paper.  However, before 
proceeding it is already possible for us to determine the degree of 
{\em relative} biasing between our galaxy types at different scales, since, 
\begin{equation} 
\frac{b^2_{\rm passive}(r)}{b^2_{\rm active}(r)} \equiv 
\frac{\xi_{\rm passive}(r)}{\xi_{\rm active}(r)} \; . 
\end{equation} 
This relative bias between our two samples is shown in Fig.~\ref{fig:xibias}, 
where we have taken the ratio between the two estimates 
of the real-space correlation function, derived in the previous 
section.  It can be seen that on small scales the clustering of the 
most passive galaxies in our sample is significantly larger than that 
of the more actively star-forming galaxies.  The relative bias then 
appears to decrease significantly until on scales greater than about 
$\sim 10\;h^{-1}$~Mpc both samples display essentially the same degree of 
clustering (within the stated uncertainties).   
 
Another frequently used method of quantifying the degree of biasing
present in a galaxy sample is through the parameter $\sigeight$ -- the
dimensionless standard deviation of (in this case) counts of galaxies
in spheres of $8\;h^{-1}$~Mpc radius.  This quantity was deemed
particularly useful because of the recognition (Peebles 1980) that for
optically selected galaxies $\sigeight\sim 1$, making the
interpretation of $b$ particularly simple since,
\begin{equation} 
b = \frac{\sigeight(\rm galaxies)}{\sigeight(\rm mass)} \;. 
\end{equation} 
Note that we write explicitly $\sigeight$ to emphasise that this is a
quantity defined on the nonlinear density field. It is an unfortunate
standard convention that, in the context of CDM models, $\sigma_8$ is
used to denote an amplitude calculated according to linear
theory. This is not the quantity considered here.
 
With $b$ defined in this way, the relative bias between our two 
spectral types is,  
\begin{equation} 
\frac{b_{\rm passive}}{b_{\rm active}} =  
\frac{\sigeight(\rm passive)}{\sigeight(\rm active)} \;. 
\end{equation} 
The quantity $\sigeight$ can be directly derived from our measured
correlation functions in quite a straight-forward manner, since the
expected variance of the galaxy counts in a randomly placed sphere is,
\begin{equation} 
\langle (N-\bar{N})^2 \rangle_R = \bar{N} + \left(\frac{\bar{N}}{V}\right)^2 \int_R 
dV_1 dV_2 \xi({\bf{r}}) \;. 
\end{equation} 
The first term is the shot-noise contribution and depends on how
sparsely we have sampled the galaxy distribution.
If we assume a power law form for the real-space correlation function
we can estimate the fluctuation amplitude with the shot noise removed
as,
\begin{equation} 
(\sigeight)^2 \equiv 
J_2(\gamma)\left( \frac{r_0}{8} \right)^{\gamma}\;, 
\end{equation} 
where, 
\begin{equation} 
J_2(\gamma) = \frac{72}{[(3-\gamma)(4-\gamma)(6-\gamma)2^\gamma]} \;, 
\end{equation} 
(Peebles 1980). 
 
Our derived values of $\sigeight$, for each of the galaxy samples
considered in the previous Section, are given in Table~\ref{tabXi}.
It can be seen from this table that the relative bias of passive with
respect to active galaxies (integrated over scales up to
$8\;h^{-1}$~Mpc) is $b_{\rm passive}/b_{\rm active}=1.09/0.75 =
1.45\pm0.14$.

\subsection{Modelling $\xi(\sigma,\pi)$} 
\label{section:modxisp}

There is much further information to be derived from the observed
$\xi(\sigma,\pi)$ grids (Fig.~1) for each galaxy type.
However, in order to do so we must first assume some model for the
clustering of galaxies with which to contrast the observations.  Here
we follow the analysis presented in Hawkins \etal\ (2003) with only
minor modifications.  Because the most significant limitation to this
analysis is inevitably the assumptions that must be enforced upon our
model, we summarise here the most important aspects of this model.
 
In order to derive a model to fit the observed $\xi(\sigma,\pi)$ grid, we 
need three main ingredients.  The first is to assume some form for the 
real-space correlation function $\xi(r)$.  Because we are only going to  
be concerned with relatively small-scale separations between galaxies  
($\le 20\;h^{-1}$Mpc), we shall assume a power-law form 
of this function,  
\begin{equation} 
\xi(r) = \left( \frac{r}{r_0} \right)^{-\gamma} \;.
\end{equation} 

In converting from real space to redshift space the next step is to
account for the distortions in the correlation function which are
caused by the linear coherent in-fall of galaxies into cluster
over-densities (Kaiser 1987; Hamilton 1992), combined with non-linear
velocity dispersion (e.g. Peacock et al. 2001).  The linear-theory
in-fall distortion can be written as (Hamilton 1992):
\begin{equation}
\xi'(\sigma, \pi) = \xi_0(s)P_0(\ct) + \xi_2(s)P_2(\ct) + \xi_4(s)P_4(\ct)\;,
\end{equation}
where P$_{\ell}(\mu)$ are Legendre polynomials, $\ct = \cos(\theta)$
and $\theta$ is the angle between $r$ and $\pi$. The relations between
$\xi_\ell$, $\xir$ and $\beta$ for a simple power-law
$\xir=(r/r_0)^{-\gamma}$ are
\begin{equation}
\label{e:xi0}
\xi_0(s) = \left(1 + \frac{2\beta}{3} + \frac{\beta^2}{5}\right)\xir\;,
\end{equation}
\begin{equation}
\xi_2(s) = \left(\frac{4\beta}{3} +
\frac{4\beta^2}{7}\right)\left(\frac{\gamma}{\gamma - 3}\right)\xir\;,
\end{equation}
\begin{equation}
\xi_4(s) =
\frac{8\beta^2}{35}\left(\frac{\gamma(2+\gamma)}{(3-\gamma)(5-\gamma)}\right)\xir
.
\label{e:xi4}
\end{equation}
We use these relations to create a model $\xi'(\sigma, \pi)$ which we
then convolve with the distribution function of random pairwise
motions, $f(v)$, to give the final model (Peebles 1980):
\begin{equation}
\xisp = \int^{\infty}_{-\infty}\xi'(\sigma, \pi - v/H_0)f(v)dv
\end{equation} 
and we choose to represent the random motions by an exponential form,
\begin{equation}
f(v) = \frac{1}{a\sqrt{2}}\exp\left(-\frac{\sqrt{2}|v|}{a}\right)
\label{e:fv}
\end{equation} 
where $a$ is the pairwise peculiar velocity dispersion (often known as
$\sigma_{12}$). An exponential form for the random motions has been
found to fit the observed data better than other functional forms
(\eg\ Ratcliffe \etal~1998; Landy 2002).

The factor $\beta \equiv \Omega_{\rm m}^{0.6}/b$ arises   
from the growth rate in linear theory,
\begin{equation} 
f \equiv \frac{d\ln\delta}{d\ln a} \approx \Omega_{\rm m}^{0.6}\;, 
\end{equation} 
which is almost independent of the cosmological constant (Lahav \etal\
1991), combined with the scale-independent biasing parameter $b$.  A
simple consequence of this model is that the redshift-space power
spectrum will also appear to be amplified compared to its real-space
counterpart\footnote{More precisely, the redshift-space distortion
factor, $\beta$, depends on the auto power spectra $P_{\rm mm}(k)$ and
$P_{\rm gg}(k)$ for the mass and the galaxies, and on the
mass-galaxies cross power spectrum $P_{\rm mg}(k)$ (Dekel \& Lahav
1999; Pen 1998; Tegmark \etal\ 2001).  The model presented here is
only valid for a scale-independent bias factor $b$ that obeys $P_{\rm
gg}(k) = bP_{\rm mg}(k) = b^2P_{\rm mm}(k)$.}. It is worth explicitly
re-stating that all of these derivations are based upon the
assumptions of the linear theory of perturbations and linear bias, and
assume the far-field approximation (although this is not a significant
issue for the 2dFGRS, for which $z_{\rm median} = 0.1$).
 
It is also interesting to note that there is another cosmological
effect that can result in the flattening of the observed
$\xi(\sigma,\pi)$ contours.  It was first noted by Alcock \& Paczynski
(1979) that the presence of a significant cosmological constant,
$\Lambda$, would result in geometric distortions of the inferred
clustering if an incorrect geometry was assumed.  However, Ballinger,
Peacock \& Heavens (1996) have shown that this is likely to be
negligible for the low redshift data-set being considered here, for
our assumed model of $\Omega_{\rm m}= 1-\Omega_\Lambda=0.3$.

To summarise, the parameters of our model are therefore the real-space
correlation function, $\xi(r)=(r/r_0)^{-\gamma}$, $\beta$, and $f(v)$,
parameterised in terms of the velocity dispersion $a$.  The
best-fitting parameters of this model can now be determined from the
model correlation function that best matches the observed
$\xi(\sigma,\pi)$.

\section{Results} 
        
\begin{table*}
 \begin{center} 
 \caption{The best fitting model parameters derived from the observed 
 $\xi(\sigma,\pi)$ grid are shown. All fits have been made over the 
 {\em quasi}-linear redshift-space separation 
 range $8 < s < 20\;h^{-1}$~Mpc.  The quoted uncertainties correspond
 to the $1\sigma$ scatter derived from the
 bootstrap estimates (see Fig.~\ref{fig:like}).  
 For comparison, uncertainties
 in $\beta$ have also been estimated from sparsely sampled 2dFGRS mock galaxy 
 catalogues (Cole \etal, 1998) limited to $z<0.15$, and correspond to
 $\Delta\beta=0.12$ for the full sample and $\Delta\beta=0.16$ for
 a 1-in-2 random sampling. This demonstrates that the
 bootstrap approach has given a fair assessment of the cosmic scatter
 in these estimates.}
 \begin{tabular}{@{}clll} 
      \hline 
   Parameter & All galaxies & Passive galaxies & Active galaxies \\ 
      \hline 
    $\beta$  &  $0.46\pm0.10$  & $0.48\pm0.14$  &  $0.49\pm0.13$ \\ 
    $a$ &  $537\pm87$ km s$^{-1}$   &$612\pm92$ km s$^{-1}$   & 
   $416\pm76$ km s$^{-1}$ \\ 
    $r_0$    & $5.47\pm0.32$ $h^{-1}$~Mpc & $7.21\pm0.34$ $h^{-1}$~Mpc 
   &  $4.24\pm0.41$ $h^{-1}$~Mpc \\  
    $\gamma$ & $1.75\pm0.08$  & $1.91\pm0.10$   &  $1.60\pm0.11$ \\ 
      \hline  
 \end{tabular} 
 \label{table:res} 
\end{center} 
\end{table*} 
 
\subsection{Validation of assumptions} 
 
We have calculated the real-space correlation function, $\xi(r)$,
independently using the non-parametric method of Saunders,
Rowan-Robinson \& Lawrence (1992), in order to confirm the range over
which it is a power-law (see Fig.~\ref{fig:xx}).  For both samples of
galaxies, $\xi(r)$ is adequately fit by a power-law to separations of
$r<20$~$h^{-1}$~Mpc.  This limit provides the upper bound to which we
can compare the observed $\xi(\sigma,\pi)$ with our assumed model.  In
addition, because we have assumed the linear theory of perturbations
in deriving our model we must impose a lower limit to the separations
that we will use in our fit.  To ensure that we have a sufficiently
large fitting range we set this lower-limit to $s=8\;h^{-1}$~Mpc.  In
fact it is quite plausible that the assumptions of linear theory are
no longer valid at this separation.  However, as will be shown, the
ability of our model to recover the observed $\xi(\sigma,\pi)$ at
these scales is reassuring.
 
On large scales it is known that the correlation function must deviate
from a pure power law form, and there is some evidence to support this
in Section~\ref{section:xir}, on scales $\sim20\;h^{-1}$~Mpc.  A
number of methods to account for this expected curvature in $\xi(r)$
were investigated in Hawkins \etal\ (2003).  However, the analysis
presented there suggests that so long as we restrict our fitting range
to $r<20\;h^{-1}$~Mpc the curvature has a negligible impact upon our
parameter estimation.  For this reason we neglect the possibility of
curvature in the present analysis and restrict ourselves to using the
simple power-law form for $\xi(r)$.
 
The other major assumption we have made is that the peculiar velocity
distribution, $f(v)$, has an exponential form.  This can be tested
using the method outlined by Landy, Szalay \& Broadhurst (1998).  This
method makes a non-parametric estimate of the velocity distribution,
using the Fourier decompositions of the observed $\xi(\sigma,\pi)$
grid along the $k_\sigma=0$ and $k_\pi=0$ axes.  Unfortunately this
method ignores the effects of coherent in-fall, which can substantially
change the resulting estimate of $a$ (see Hawkins \etal\ 2003).
However, it is found that the recovered $f(v)$ is well fit by an
exponential form -- for both types of galaxies.  Hawkins \etal\ (2003)
have shown that although incorporating the effects of coherent in-fall
changes the estimated velocity dispersion, $a$, substantially, the
method gives a robust estimate of the form for $f(v)$.

\subsection{Parameter fits} 
 
All four of our parameters ($\beta$, $a$, $r_0$ and $\gamma$) are
allowed to vary over a large range of possibilities and a {\em
downhill simplex} multi-dimensional minimisation routine is adopted to
find their best-fitting values (see e.g. Press \etal\ 1992).  Our
calculated $\xi(\sigma,\pi)$ contours are shown in
Fig.~\ref{fig:plate}, together with those of the best-fitting model
correlation functions derived in this manner.  The peak parameters of
this best-fitting model are detailed in Table.~\ref{table:res}
together with their estimated uncertainties.  We find that there is
quite a significant degeneracy between $\beta$ and $a$ (see
Fig.~\ref{fig:like}).  This is also exacerbated by the relatively noisy
nature of $\xi(r)$ at these scales, which makes $r_0$ and $\gamma$
difficult to constrain accurately.
 
One immediate conclusion is that the velocity dispersions of the two
galaxy populations are very distinct, even taking into account the
substantial statistical uncertainties.  This is an interesting result
which has significant implications for the proportion of each of these
galaxy types we expect to occupy large, virialised clusters of
galaxies.
 
Another conclusion that we can easily make is to quantify the relative
bias between our two spectral types, as described in
Section~\ref{section:relbias}.  However, as demonstrated in that
Section, the relative bias between our galaxy types is in fact
essentially unity over the range for which our model assumptions are
valid ($8-20\;h^{-1}$~Mpc), a result confirmed in this analysis.
However, a much more important quantity that can be inferred from
these redshift-space distortions, that could not be determined
previously, is the absolute value of the biasing
between the galaxy and mass distributions, $b$, as described in
Section~\ref{section:relbias}.  We return to this point in the next
Section of this paper.

\begin{figure*} 
\psfig{figure=xieta_fig5a.ps,angle=-90,width=3.4in}
\psfig{figure=xieta_fig5b.ps,angle=-90,width=3.4in} 
\caption{The full $\xi(\sigma,\pi)$ grids for our different spectral
types: passive (left) and active (right). Also plotted (white lines)
are the contour levels of the best-fitting model derived earlier.  The
contour levels are $\xi$ = 4.0, 2.0, 1.0, 0.5, 0.2, 0.1.}
\label{fig:plate}
\end{figure*}

\begin{figure} 
\begin{center} 
\psfig{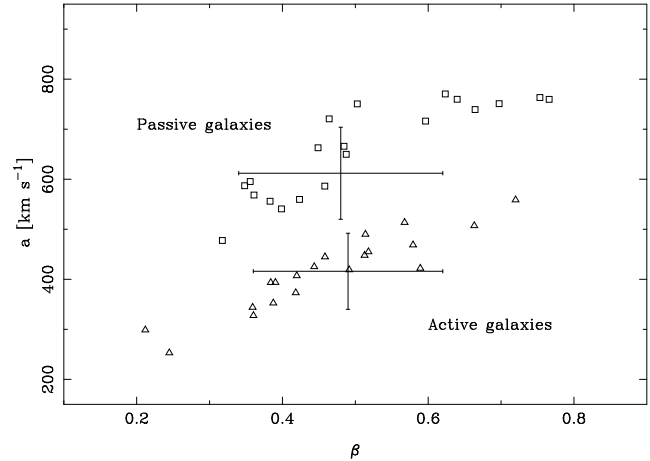}
\caption{The estimates of $\beta$ and $a$ for the bootstrap samples are
shown for both the passive (squares) and active (triangles) galaxy
samples.  It can be seen that a significant degeneracy exists between
$\beta$ and $a$, for both samples. The fits
have been made using only the {\em quasi}-linear regime of
$8-20\;h^{-1}$Mpc.  The crosses show the fits to the full samples
together with the $1\sigma$ uncertainties shown in Table~2.}
\label{fig:like} 
\end{center} 
\end{figure}

\section{Comparison with Previous Results} 
\label{section:compxi} 

\begin{table*}
 \begin{center} 
 \caption[Comparison between biasing results derived using the 2dF Galaxy 
 Redshift Survey by various authors.] 
 {Comparison between biasing results derived using the 2dF Galaxy 
 Redshift Survey by various authors. The results of Peacock \etal\ 
 (2001) are derived from the redshift-space distortions in the 
 two-point correlation function (Note that this galaxy sample was much 
 deeper as galaxies without spectral types were used). Lahav \etal\ 
 (2002) made their estimate of the bias through comparing the 
 amplitude of fluctuations in both the 2dFGRS and the CMB.  Verde 
 \etal\ (2002) calculated the bi-spectrum of the 2dFGRS, which 
 constrained the linear bias parameter, $b$, which we have converted to 
 $\beta$ by assuming our concordance cosmological model of a flat 
 Universe with $\Omega_{\rm m}=0.3$.  Note that the results of Lahav 
 \etal\ (2002) and Verde \etal\ (2002) are valid over scales expressed 
 in terms of wavenumber $k$ rather than real-space distance.  We have 
 converted between the two by simply taking $r\sim1/k$.} 
 \begin{tabular}{@{}cccccc@{}} 
 \hline 
 Galaxy type & Author & Scales ($h^{-1}$Mpc) & ($z_{\rm s}$, $L_{\rm s}/L_*$) & 
 $\beta(z_{\rm s},L_{\rm s})$ & $\beta(0,L_*)$ \\ 
 \hline 
  All & -- & $8-20$ & (0.11,1.06) & $0.46\pm0.10$ & 
 $0.44\pm0.09$ \\ 
  Passive & -- & $8-20$ & (0.11,1.26) & $0.48\pm0.14$ & 
 $0.47\pm0.14$  \\ 
  Active & -- & $8-20$ & (0.11,0.95) & $0.49\pm0.13$ & 
 $0.48\pm0.13$  \\ 
\hline 
  All & Hawkins \etal\ & $8-20$ & (0.15,1.4) & $0.49\pm0.09$   &  $0.47\pm0.08$ \\ 
  All & Peacock \etal\ & $8-25$ & (0.17,1.9) & $0.43\pm0.07$ & $0.45\pm0.07$ \\ 
  All & Lahav \etal\ & $7-50$ & (0.17,1.9)   & $0.48\pm0.06$ & $0.50\pm0.06$ \\ 
  All & Verde \etal\ & $2-10$ & (0.17,1.9) & $0.56\pm0.06$ & $0.59\pm0.06$ \\ 
\hline 
\end{tabular} 
\label{tab:xires} 
\end{center} 
\end{table*}

Because previous estimates of $\beta$ have used slightly different
galaxy samples, it is first necessary to correct for various effects
before making a proper comparison (see Lahav \etal\ 2002).  There are
two main issues which affect the different estimates of the biasing;
the effective redshift of the survey sample used, $z_{\rm s}$, and the
effective luminosity of the galaxies in that sample, $L_{\rm s}$.  These
quantities vary between the samples used, depending on the weighting
scheme adopted and the limiting redshift of the survey.
 
Because we have only used 2dFGRS galaxies for which a spectral type is 
available our sample is limited to $z_{\rm max}=0.15$.  To determine 
the effective redshift of our sample it is necessary for us to determine 
the {\em weighted} average of the galaxies used in each of our 
calculations.  Doing so reveals 
that for all three of our samples $z_{\rm s} = 0.11$.  In a similar way 
we can calculate the weighted mean luminosity of each of our samples, 
which are found to be as follows. 
Combined:  $L_{\rm s} = 1.06 L_*$; 
Passive:   $L_{\rm s} = 1.26 L_*$; 
Active:    $L_{\rm s} = 0.95 L_*$; 
where we have taken $M^*-5\log_{10}(h) = -19.66$ (Norberg \etal\ 
2002b). 
 
Assuming linear dynamics and linear biasing, the redshift-distortion
parameter, $\beta$, for a given sample redshift and luminosity can be
written as,
\begin{equation} 
\beta(L,z) \approx \frac{\Omega_{\rm m}^{0.6}(z)}{b(L,z)} \;. 
\end{equation} 
The evolution of the matter density parameter, $\Omega_{\rm m}(z)$ is 
straight-forward to determine, assuming a given cosmological model, 
\begin{equation} 
\Omega_{\rm m}(z) = \Omega_{\rm m} (1+z)^3 (H/H_0)^{-2} \;, 
\end{equation} 
where $\Omega_{\rm m}$ is the matter density at the present epoch and, 
\begin{equation} 
\left( \frac{H}{H_0} \right)^2 = \Omega_{\rm m}(1+z)^3 + 
(1-\Omega_{\rm m}-\Omega_\Lambda)(1+z)^2 + \Omega_\Lambda \;. 
\end{equation} 
 
The determination of the variation in the biasing parameter, $b$, with
redshift is much less straight-forward.  As shown in
Section~\ref{section:relbias}, $b$ can be defined as,
\begin{equation} 
b(z) = \frac{\sigeight{}^{,\, \rm g}(z)}{\sigeight{}^{,\, \rm m}(z)} \;, 
\end{equation} 
where here we have added a redshift dependence, $b(z)$, and labelled
the two $\sigeight$'s by the superscripts $\rm g$ and $\rm m$ to
denote galaxies and mass respectively.  As described by Lahav \etal\
(2002), there is now much evidence to suggest that whilst the matter
fluctuations continue to grow at low redshifts, the fluctuations in
the galaxy distribution are relatively constant between $0 < z < 0.5$
(see e.g. Shepherd \etal\ 2001).  If we assume that the matter
fluctuations grow according to the linear theory of perturbations
then, $\sigeight{}^{,\, \rm m}(z) = \sigeight{}^{,\, \rm m}(0) D(z)$,
where $D(z)$ is the growing mode of fluctuations (Peebles 1980).
Whereas, $\sigeight{}^{,\, \rm g}(L,z) \approx \sigeight{}^{,\, \rm
g}(L,0)$.  Therefore,
\begin{equation} 
b(L,z) = \frac{b(L,0)}{D(z)} 
\end{equation} 
The final step then is to correct the biasing parameter, $b$, for the
luminosity of our sample.  Norberg \etal\ (2001) found from the
analysis of the galaxy correlation functions on scales $<10\;h^{-1}$
Mpc that,
\begin{equation} 
\frac{b(L,0)}{b(L_*,0)} = 0.85 + 0.15\left( \frac{L}{L_*} \right) \;. 
\end{equation} 
Assuming that this relation also holds in our quasi-linear regime of
$8-20\;h^{-1}$~Mpc, then allows us to determine $\beta$ at redshift
$z=0$ and luminosity $L=L_*$.\footnote{Note that in converting the
linear bias parameter, $b(L,0)$, for each of our spectral types to
$b(L_*,0)$, we have explicitly assumed that each type of galaxy
displays the same variations in clustering with luminosity.  This
result has been verified by Norberg \etal\ (2002a), who calculated the
clustering amplitudes for different galaxy samples divided in spectral
type and luminosity.}
 
Table~\ref{tab:xires} shows the results for $\beta$ derived in the
analysis presented here, both before and after converting to redshift
$z=0$ and luminosity $L=L_*$.  Also shown are other results derived
from the 2dFGRS by previous authors.  It can be seen that there is a
remarkably good agreement between all the results presented.  We note
that these results have been derived by applying {\em linear}
corrections to a selection of {\em quasi-linear} regimes, which may
introduce systematic errors into our results.  This is a particular
concern for the results of Verde \etal\ (2002), which correspond to
the smallest separation ranges used.

\section{Discussion} 
\label{section:conclusion} 
 
We have derived a variety of different parameterisations for the
2dFGRS correlation function, $\xi(\sigma,\pi)$, for different spectral
types.  The two types we have used can roughly be interpreted as
dividing our galaxy sample on the basis of their relative amount of
current star-formation activity, and hence provide useful insight into
how galaxy formation may relate to the large-scale structure of the
galaxy distribution.  The actual cut we have imposed is most naturally
interpreted in terms of the Scalo birthrate parameter, $b_{\rm
Scalo}$. This is defined to be the ratio of the current star-formation
rate and the past averaged star-formation rate.  Adopting this
convention, our cut of $\eta=-1.4$ corresponds to dividing our sample
into galaxies with $b_{\rm Scalo}=0.1$, i.e. between galaxies whose
present star formation rate is greater or less than 10\% of their past
averaged rate.
 
\subsection{Relative bias on small scales}  
 
On scales smaller than $\sim8\;h^{-1}$~Mpc the clustering of passive
galaxies is much stronger than that of the more actively star-forming
galaxies.  This was demonstrated quantitatively by the real-space
correlation functions derived in Section~\ref{section:xir}, for which
the passive galaxy sample were fit by a power-law with larger scale
length, $r_0$ and steeper $\gamma$.  In addition it was shown that the
values of $\sigeight$ derived for each of these samples were quite
distinct, being $\sigeight=1.09\pm0.08$ for the passive galaxies and
$\sigeight=0.75\pm0.09$ for the actively star-forming galaxies,
implying an (integrated) relative bias between our two types of,
\begin{equation} 
\frac{b_{\rm passive}}{b_{\rm active}} =  1.45\pm0.14 \;,  
\end{equation} 
at the effective redshift and luminosity of our galaxy samples (see
Table~3).  Note that this ratio quantifies the {\em integrated}
relative bias between scales of $0-8\;h^{-1}$~Mpc.

Our correlation functions per type confirm that the slope of passive
(early type) galaxies is steeper than that of active (late tape)
galaxies (cf. Zehavi et al. 2001).  On the other hand, the slope of
the correlation functions derived for different luminosity ranges show
no significant variation (Norberg et al. 2001, 2002a; Zehavi et
al. 2001).  These results call for theoretical explanations and they
set important constraints on models for galaxy formation.

\subsection{Velocity dispersions} 
 
The velocity distributions of our two samples were found to be
distinct.  The passive galaxy sample displayed a consistently larger
velocity dispersion, $a$, than the actively star-forming sample on all
scales, and in particular on separations of $8-20\;h^{-1}$~Mpc were
found to be $612\pm92$ and $416\pm76$ km s$^{-1}$ respectively.
This result is consistent with the observations of Dressler (1980),
that a significant morphology-density relation exists -- since a
larger velocity dispersion would tend to suggest a higher proportion
of galaxies occupying virialised (high-density) clusters.

\subsection{Relative bias on large scales}  
 
The determination of the redshift-distortion parameter, $\beta$, was
found to be much less straight-forward.  The evidence from our
analysis is that $\beta$ has only a relatively small dependence on the
spectral type of the galaxy sample under investigation.  We found that
on scales of $8-20\;h^{-1}$~Mpc, the two redshift distortion
parameters were; $\beta=0.48\pm0.14$ and $\beta=0.49\pm0.13$ for the
passive and actively star-forming galaxy samples respectively,
yielding a relative bias of only,
\begin{equation}
 \frac{b_{\rm passive}}{b_{\rm active}}=1.02\pm0.40.
\end{equation} 
The overall redshift-distortion parameter, $\beta$, independent of
spectral type is found here to be $0.46\pm0.10$ (on scales of
$8-20\;h^{-1}$~Mpc), at our sample's mean redshift and luminosity.  By
making various assumptions (Section~\ref{section:compxi}) this result
can be converted to redshift $z=0$ and $L_*$ luminosity, giving
$\beta(0)=0.44\pm0.09$.  This result is is almost identical to the
$\beta(0)=0.47\pm0.08$ derived from the results of Hawkins \etal\ (2003),
using the entire 2dFGRS data-set, over the same separation range.
 
In the analyses presented in this paper two fundamental limits were
found to greatly inhibit our ability to accurately characterise the
relative and absolute biases on different scales.  The first of these
was that on small scales -- where the clustering of our two
populations is most distinct -- the assumptions of our model of the
galaxy clustering were no longer accurate, and so we could not
accurately determine $\beta$ or $a$ on these scales.  Our second
limitation was found on large scales ($s\sim20\hmpc$), where the
galaxy correlation functions became noisy and were no longer well
parameterised by a power-law form.
 
The latter of these issues can be addressed to some degree simply by a
change of formalism to incorporate the power spectrum estimations of
each galaxy type or colour (Peacock 2003).  Because this
characterisation of the clustering is more sensitive to larger scales
of separations it would allow us to more rigorously test whether the
large-scale ($s > 20\hmpc$) clustering of these populations are in
fact distinct and also allow us to incorporate the possibility of
scale-dependent bias.  The derived correlation functions per type
could also be used within the framework of halo occupation number to
derive e.g. the mean number of galaxies of a given type per halo
(e.g. Zehavi et al. 2003; Magliochetti \& Porciani 2003).

\section*{Acknowledgements} 
 
DSM was supported by an Isaac Newton Studentship from the Institute of 
Astronomy and Trinity College, Cambridge.  The 2dF Galaxy Redshift 
Survey was made possible through the dedicated efforts of the staff at 
the Anglo-Australian Observatory, both in creating the two-degree 
field instrument and supporting it on the telescope.

\bsp 
\label{lastpage} 
\end{document}